\documentclass{ifacconf}

\usepackage{amsmath,amssymb,amsfonts}
\usepackage{comment}
\usepackage{appendix}

\usepackage{graphicx}
\usepackage{textcomp}
\usepackage{xcolor}
\usepackage{tikz}
\usetikzlibrary{patterns}
\usetikzlibrary{calc,shapes,arrows}

\usepackage{algpseudocode}

\usepackage{subcaption}

\newtheorem{assumption}{Assumption}
\newtheorem{problem}{Problem}
\newtheorem{definition}{Definition}
\newtheorem{theorem}{Theorem}
\newtheorem{proposition}{Proposition}
\newtheorem{corollary}{Corollary}

\newtheorem{remark}{Remark}

\newcommand{\R}{{\mathbb{R}}}

\newcommand{\B}{{\mathbb B}}

\newcommand{\N}{{\mathbb{N}}}

\newcommand{\X}{{\mathbb{X}}}
\renewcommand{\S}{{\mathbb{S}}}

\newcommand{\Y}{{\mathbb{Y}}}

\usepackage{bm}
\newcommand{\Hilbert}{\mathcal{H}}
\newcommand{\innerH}[3]{\langle #1, #2 \rangle_{#3}\!} 
\newcommand{\T}{^\top}
\newcommand{\cdotx}{\,\cdot\,}
\newcommand{\map}{\Psi}
\newcommand{\mapq}{\Psi_q}

\newcommand{\Tr}{\mathbf{t}}

\newcommand{\M}{\mathbf{M}}
\newcommand{\borel}[1]{\mathcal{B}(#1)}

\newcommand{\p}{\mathbf{p}}
\newcommand{\trace}[1]{\mathrm{tr}(#1)}

\newcommand{\1}{\mathbf{1}}

\newcommand{\data}{\mathcal{D}_N}
\renewcommand{\P}{\mathbb{P}}

\newcommand{\target}{\mathcal{T}}

\def\optimize{\textsf{optimize}}
\usepackage{url}
\usepackage{fancyhdr}
\newcommand{\red}{\textcolor{red}}

\definecolor{TACcyan}{RGB}{0,141,225}
\colorlet{accent}{TACcyan}
\definecolor{brightmagenta}{rgb}{1, 0, .5}
\colorlet{accentwashed}{TACcyan!15}
\definecolor{grayfilling}{gray}{0.95} 
\definecolor{grayshadow}{gray}{0.5} 

\usepackage{algorithm}
\usepackage{graphicx}      
\usepackage{natbib}        

\newif\iflong 
\longtrue

\begin{document}
\begin{frontmatter}

\title{Data-Driven Abstractions via Binary-Tree Gaussian Processes for Formal Verification}


\author[First]{Oliver Sch\"on} 
\author[First]{Shammakh Naseer} 
\author[First]{Ben Wooding}
\author[Second]{Sadegh Soudjani}

\address[First]{School of Computing, Newcastle University, UK}
\address[Second]{Max Planck Institute for Software Systems, Germany}

\begin{abstract}                
To advance formal verification of stochastic systems against temporal logic requirements for handling unknown dynamics, researchers have been designing data-driven approaches inspired by breakthroughs in the underlying machine learning techniques.
As one promising research direction, abstraction-based solutions based on Gaussian process (GP) regression have become popular for their ability to learn a representation of the latent system from data with a quantified error. Results obtained based on this model are then translated to the true system via various methods.
In a recent publication, GPs using a so-called binary-tree kernel have demonstrated a polynomial speedup w.r.t. the size of the data compared to their vanilla version, outcompeting all existing sparse GP approximations.
Incidentally, the resulting binary-tree Gaussian process (BTGP) is characteristic for its piecewise-constant posterior mean and covariance functions, naturally abstracting the input space into discrete partitions.
In this paper, we leverage this natural abstraction of the BTGP for formal verification, eliminating the need for cumbersome abstraction and error quantification procedures.
We show that the BTGP allows us to construct an interval Markov chain model of the unknown system with a speedup that is polynomial w.r.t. the size of the abstraction compared to alternative approaches.
We provide a delocalized error quantification via a unified formula even when the true dynamics do not live in the function space of the BTGP.
This allows us to compute upper and lower bounds on the probability of satisfying reachability specifications that are robust to both aleatoric and epistemic uncertainties.
\end{abstract}


\end{frontmatter}

\section{Introduction}
Developing the control software for safety-critical cyber-physical systems such as power grids, autonomous vehicles, medical implants, and robotic systems has raised the need for formal approaches to guarantee the systems will behave as expected, and not lead to catastrophe. 
Designing controllers for such systems is especially challenging for their stochastic nature.
Previous approaches were mainly based on physics-based models of the underlying systems. To meet the increasing need for more scalable and uncertainty-robust techniques, data-driven approaches govern the contemporary landscape of active research.
One powerful offspring that is being explored is \emph{Gaussian process} (GP) regression~\citep{Rasmussen2018GP}. GPs use a Bayesian paradigm to learning unknown functions from observations, that is particularly valuable for safety-critical applications as the statistical approximation error is naturally quantified.

In this work, we consider GPs based on the so-called \emph{binary-tree} (BT) \emph{kernel} \citep{cohen2022log}, a discontinuous kernel
that has been shown to reduce the computational burden of GP regression significantly, increasing its scalability to higher dimensional systems and larger data sets.
In comparison to alternative approaches used for constructing scalable GPs (see, e.g., \cite{quinonero2005unifyingGP}), the BT kernel achieves GP regression in linear time -- a polynomial improvement.
The resulting \emph{binary-tree Gaussian process} (BTGP) is characteristic for its piecewise-constant posterior mean and covariance functions, resulting in a natural partitioning of its input space.
The main idea of this paper is to exploit this direct learning of an abstract representation of the unknown data-generating system for formal verification.

Formal verification of known stochastic systems against temporal logic properties has been well studied \citep{Baier2008TempLogics}.
For settings where the system dynamics are uncertain -- termed \emph{epistemic uncertainty} --- and specifications have infinite horizons, most available abstraction-based approaches can be divided into two categories.
Approaches based on statistical relations replace the original system with an estimated surrogate model that does not capture the epistemic uncertainty, generally a variant of a \emph{Markov chain} (MC). Results obtained on this model hence rely on quantifying the probabilistic deviation to the true system to transfer guarantees \citep{abate2008markov,haesaert2018POMDP}.
For example, \emph{probabilistic coupling relations} can be used to transfer guarantees obtained on a parameterized surrogate model of the unknown system back to the original latent system by quantifying the expected \emph{parametric uncertainty} \citep{schon2023bayesian}. 
Verification is conducted using a robust version of dynamic programming \citep{haesaert2020robust}.
Establishing such coupling relations, however, requires the computation of local error parameters quantifying the error between the true system and its surrogate model for every discrete partition, deeming their application to high-dimensional systems computationally excruciating.
In an effort to avoid the construction of surrogate models requiring such types of relations, approaches based on \emph{interval Markov chains} (IMCs)
incorporate both stochasticity and epistemic uncertainty \citep{Badings2022epistemicUncert, Lavaei2022DDMDP,Jackson2020safety}.
The IMCs
constructed based on upper and lower bounds on the transition probabilities, learned, e.g., using GP regression \citep{Jackson2020safety}, can be subsequently verified via the \emph{interval iteration algorithm} \citep{haddad2018interval}.
Whilst this alleviates the need for constructing additional coupling relations, the computations required to construct and verify the IMC
are substantially more involving than the corresponding computations on the MC
in the coupling-based approach.
%

In this paper, we make an effort to overcome this compromise.
We use BTGP regression to learn the unknown dynamics of a discrete-time continuous-space stochastic system and show that the probability bounds of the resulting IMC are considerably easier to compute by obtaining transition probability distributions from the BTGP that are naturally piecewise-constant.
Whilst prior IMC-based approaches require the optimization of both probability bounds to identify the worst-case probabilities for every transition, this change allows us to perform only two integrations for each transition -- one for either bound -- and obtain equivalent robustness results. Hence, we get a speedup for the construction of the abstraction that is polynomial w.r.t. the size of the abstraction.
With this, we make the following contributions.
\begin{itemize}
    \item We provide a data-driven approach to verify uncertain stochastic systems defined over continuous spaces against infinite-horizon reachability specifications based on an IMC generated via BTGP regression. In comparison to existing approaches, the IMC is constructed without the need for the identification of the worst-case bounds for every partition. This yields a computational speedup over IMC abstraction that is polynomial in the number of partitions.
    \item We obtain statistical guarantees on the accuracy of the BTGP model that combine available measure-theoretic results with an error inflicted by the approximation error of the BTGP even if the true dynamics function does not live in the BT kernels RKHS.
    \item We show how a discontinuous kernel can be used to generate a finite abstraction of the latent model and how this abstraction can be efficiently verified.
    \item We give closed-form expressions for the function space generated by the BT kernel.
\end{itemize}

\smallskip

\noindent\textbf{Related work:}
We give a brief review of relevant related approaches on data-driven formal verification and synthesis of uncertain stochastic systems.

%
The works by \cite{Badings2022epistemicUncert} and \cite{Lavaei2022DDMDP} address epistemic uncertainty in stochastic systems by abstracting the system to an interval \emph{Markov decision process} (MDP).
While the prior is limited to finite horizons and linear systems, the latter requires solving a scenario optimization problem.
\cite{Jackson2020safety} and \cite{jiang2022safe} have used GPs instead of scenario optimization; see also the references therein for more akin literature.
Non-parametric estimation and abstraction to interval MDPs for formal verification is studied by \cite{zhang2024formal}.
In a related kernel-based setting, \cite{thorpe2022cmechance} use conditional mean embeddings (CMEs) to embed the conditional distribution of a non-Markovian random trajectory into an RKHS. 
%
Temporal logic control of systems captured by stochastic neural network dynamic models (NNDM) is addressed by \cite{adams2022nndm} by translating the NNDMs to IMDPs.
The starting point, however, is the NNDM and no guarantees on the correctness of the learned model w.r.t. the data-generating system are considered, rendering the approach only robust from a model-based perspective.
%
Capturing epistemic uncertainty using interval MDPs relies on the assumption of state-wise independent epistemic uncertainty, leading to inconsistent and overly conservative results.
In contrast, \cite{schon2023bayesian} capture epistemic uncertainty explicitly using parametric MDPs and probabilistic coupling relations to the latent true system. Furthermore, their results are applicable to a wider range of specifications and distributions.
%
The neural abstractions studied by \cite{abate2022neuralabstract} rely on a counterexample-producing scheme via SMT solvers.
%
A contraction-guided adaptive partitioning approach is introduced by \cite{harapanahalli2023contraction}.
The authors of \cite{kazemi2024data} compute a growth bound of the unknown system and design controllers founded on estimating a bound on the Lipschitz constant of the system.

Latest work in the area of abstraction-free approaches is based on constructing a set of feasible models from data and finding a common control barrier certificate (CBC)  \citep{Cohen2022RCBF, Lopez2022UnmatchCBC}.
\cite{salamati2024data} find CBCs using robust scenario optimization. Whilst this necessitates them to sample transitions between pairs of partitions multiple times, 
they also require knowledge of the Lipschitz constants of several intermediate functions.
The work by \cite{schon2024data} overcomes these limitations through the use of CMEs.
The work by \cite{wajid2022formal} addresses the problem of finding CBCs using GPs.
Whilst CBCs are often deemed more scalable than abstraction-based solutions, finding a valid CBC for systems with non-control-affine and non-polynomial dynamics is generally hard. Even more, CBCs for specifications beyond simple reachability yields, e.g., sequential reachability problems \citep{jagtap2020},
greatly impeding their applicability.
%
Further work on model-free reinforcement learning studies synthesizing robust temporal logic controllers without constructing an explicit model of the system \citep{hasanbeig2019RL, kazemi2020fullLTL,kazemi2024assume}.
%
An approach based on computing reachable sets leverages random set theory to obtain infinite-sample guarantees is provided by \cite{lew2021sampling} without analyzing finite-sample convergence rates.
Similarly, \cite{cao2022efficient} construct reachable sets efficiently using GPs under the assumption of known bounds on the unknown dynamics components.
There exists a body of work on verifying models composed of neural networks, e.g., \citep{harapanahalli2023forward, jafarpour2023interval}. However, similar as mentioned before, no guarantees for the correctness of the data-driven model w.r.t. the data-generating system are provided.

\smallskip

The remainder of this manuscript is structured as follows. After providing preliminary definitions and the problem statement in Sec.~\ref{sec:prelim_and_prob}, we dive into system identification using BTGP regression in Sec.~\ref{sec:sysid_via_btgp}. Here, we provide our results examining the BTGP and its associated statistical error bound.
In Sec.~\ref{sec:verif}, we demonstrate how the BTGP allows us to generate an IMC model efficiently and provide a suitable verification algorithm.
In Sec.~\ref{sec:casestudies},
we showcase the presented approach on a nonlinear case study.
Due to space restrictions, the proofs of statements 
\iflong are relegated to the appendix.
\else will be provided in an extended version.
\fi

\section{Preliminaries and Problem Statement}
\label{sec:prelim_and_prob}
\vspace{-0.1cm}
Let $\R$, $\R_{\ge 0}$, $\N$, and $\B$ be the sets of reals, non-negative reals, positive integers, and binary numbers.
Let $s\in\mathbb{B}^q$ be a binary string of length $|s|:=q$. We denote its substring prefix of length $l\in\N$ by $[s]_{\leq l}$.
Let $I_N$ be the $N\times N$ identity matrix.
The transpose of a vector or matrix $A$ is indicated by $A\T$.
For some topological space $\X$, let $X_N:=[x_1,\ldots,x_N]\T$ be a column vector with $x_i\in\X$. 
To save space, we may use a compact notation and write the column vector as $X_N:=[x_i]_{i=1}^N$.
We denote the element-wise evaluation of a function $f:\X\rightarrow\Y$ on $X_N$ by $f(X_N):=[f(x_i)]_{i=1}^N$.
Similarly, we may write $A = [a_{ij}]_{i,j=1}^N$ to denote a matrix with its elements. The trace of such a matrix is denoted by $\trace{A}: = \sum_i a_{ii}$.
For $b\in\{\top,\bot\}$, the indicator function is given by $\1(b):=\{1, \text{ if } b=\top, \text{ and $0$ otherwise}\}$.

\noindent\textbf{Probability theory:}
A \emph{probability space} is a tuple $(\X,\borel{\X},p)$ equipped with a sample space $\X$, a Borel $\sigma$-algebra $\borel{\X}$ defined over $\X$, i.e., the smallest $\sigma$-algebra containing open subsets of $\X$, and a probability measure $p$, which has realizations  $x\sim p(\cdotx)$.
In this work, we restrict our attention to Polish sample spaces~\citep{bogachev2007measure}.
A \emph{probability measure} $p$ on a measurable space $(\X,\borel{\X})$ is a map
$p:\borel{\X}\rightarrow [0,1]$ such that for all countable collections $\{A_i\}_{i=1}^\infty$ of pairwise disjoint sets in $\borel{\X}$ it holds that $p({\bigcup_i A_i })=\sum _i p({A_i})$ and $p(\X)=1$.
For two measurable spaces $(\X,\borel{\X})$ and $(\Y,\borel{\Y})$, a \emph{probability kernel} is a mapping $\p: \X \times \borel{\Y}\rightarrow  [0,1]$ such that $\p(x,\cdotx):\borel{\Y}\rightarrow[0,1]$ is a probability measure for all $x\in\X$, and $\p(\cdotx, Y): \X\rightarrow [0,1]$ is measurable for all  $Y\in\borel{\Y}$.
A probability kernel associates to each point $x\in\X$ a measure, also denoted by $\p(\cdotx|x)$.
Given $n\in\N$, the (Gaussian) normal probability measure with mean $\mu\in\mathbb{R}^n$ and covariance matrix $\Sigma\in\mathbb{R}^{n\times n}$ is denoted 
as $\mathcal N(\cdotx|\mu, \Sigma)$.

\noindent\textbf{RKHS theory:}
A positive-definite, symmetric function $k:\X\times\X\rightarrow\R$ is called a \emph{kernel} (note the distinction from \emph{probability kernels}) if for all $N\in\N$, $a_1,\ldots,a_N\in\R$, and $x_1,\ldots,x_N\in\X$ we have $\sum_{i=1}^{N}\sum_{j=1}^{N}a_ia_jk(x_i,x_j)\geq0$.
Note that a positive-definite kernel gives rise to a positive-definite \emph{Gram matrix} $K:=[k(x_i,x_j)]_{i,j=1}^N$.
Given a kernel $k$ on a non-empty set $\X$, there exists a corresponding unique \emph{reproducing kernel Hilbert space} (RKHS) $\Hilbert_{k}$ of functions.
$\Hilbert_{k}$ is equipped with an inner product $\innerH{\cdotx}{\cdotx}{\Hilbert_{k}}$ with the \emph{reproducing property} such that for any function $f:\X\rightarrow\R$, $f\in\Hilbert_{k}$ and $x\in\X$ we have $f(x)=\innerH{f}{k(\cdotx,x)}{\Hilbert_{k}}$ where $k(\cdotx,x):\X\rightarrow\Hilbert_{k}$ is a real-valued function for which $k(x,x')=\innerH{k(\cdotx,x)}{k(\cdotx,x')}{\Hilbert_{k}}$ for all $x,x'\in\X$.
The inner product induces the norm $||f||_{\Hilbert_{k}}:=\sqrt{\innerH{f}{f}{\Hilbert_{k}}}$. 
A kernel $k$ is called \emph{translation-invariant} if it can be expressed as a function of distance between points $x,x'\in\X$, i.e., there exists some function $\Delta:\X\rightarrow\R$ such that $k(x,x')= \Delta(x-x')$.

\noindent\textbf{Dynamical systems:}
Consider a system $\M$ with a state space $\X\subset\R^n$ of dimension $n\in\N$ and dynamics
\begin{equation}
    \M: 
    \begin{array}{ll}
		x_{t+1}= f(x_t) + v_t,\quad v_t\sim \mathcal{N}(\cdotx|0,\sigma^2_v I_n),
	\end{array}
 \label{eq:system}
\end{equation}
where $x_t\in\X$
denotes the system state
at the time instance $t\in\N\cup\{0\}$.
The \emph{independent, identically distributed} (i.i.d.) process noise $v_t$ is zero-mean Gaussian with covariance matrix $\sigma^2_v I_n$. Note that this implies that $v_t$ is dimension-wise independent.
The transition function $f:\X\rightarrow\X$ is unknown and we assume that it lives in the RKHS of some translation-invariant kernel $k$, i.e., $f\in\Hilbert_{k}$. The covariance matrix of the process noise is known.

\noindent\textbf{Reachability specifications:}
Given a target set $\target\subset\X$, the system satisfies the reachability specification $\psi_\target$ with probability at least $p_\psi\in[0,1]$ if the probability that the trajectories remain in $\X$ until eventually reaching the target set $\target$ is greater than or equal to $p_\psi$. This is denoted by $\P(\M\models\psi_\target)\geq p_\psi$. For the purpose of this work, we limit ourselves to reachability specifications, but the provided verification algorithm can be applied to any linear temporal logic specification by adding a preprocessing step on the graph structure of the constructed abstract model \citep{Baier2008TempLogics,haddad2018interval}.

\noindent\textbf{Problem statement:}
Let $\data:=\{x_i,y_i\}_{i=1}^N\subset\X\times\X$ be a set of $N\in\N$ data samples from the latent true system in \eqref{eq:system} such that $y_i:=f(x_i)+v_i$ are output samples with i.i.d. noise $v_i\sim\mathcal{N}(\cdotx|0,\sigma^2_v I_n)$. We may write $\data=(X_N,Y_N)$, where $X_N:=[x_1,\ldots,x_N]$ and $Y_N:=[y_1,\ldots,y_N]\T$.
%

\begin{problem}
    Let data $\data$ from the system in \eqref{eq:system} with unknown function $f\in\Hilbert_{ k}$ be given.
    For an infinite-horizon reachability specification $\psi_\target$,
    verify that the system satisfies $\psi_\target$ with probability at least $p_\psi\in[0,1]$.
\end{problem}
To address this problem, we construct a function estimator $\hat f:\X\rightarrow\X$ based on $\data$ that approximates the unknown function $f\in\Hilbert_{k}$ in \eqref{eq:system}.
In particular, we specify a kernel $\hat k$ characterizing the function class of $\hat f\in\Hilbert_{\hat k}$.
To be able to address a wealth of functions $f$ and perform the function approximation in a function space $\Hilbert_{\hat k}\not\supset\Hilbert_{ k}$, we raise the following assumption.
\begin{assumption}\label{asm:knownkernel}
    Let the unknown function $f\in\Hilbert_{ k}$ and the function estimator $\hat f\in\Hilbert_{\hat k}$, with known kernels $k, \hat k:\X\times\X\rightarrow\R$
    and complexity bound $B\geq||f||_{\Hilbert_{ k}}\geq0$ .
\end{assumption}
Moving forward, we use `$\string^$' to denote symbols referring to the estimated model.
%
%
\begin{remark}
    We raise Assumption~\ref{asm:knownkernel} to allow for greater flexibility in choosing the kernel $\hat k$ used for estimation. In particular, we will use a discontinuous kernel. To still be able to address more arbitrary functions, such as continuous functions living in the RKHS of the squared exponential kernel, we need to quantify the error introduced by constructing an estimator in a different RKHS.
\end{remark}

\section{System Identification Using Binary-Tree Gaussian Processes}
\label{sec:sysid_via_btgp}
In this section, we give a brief overview of the classical GP regression framework, followed by an examination of GPs based on the binary-tree kernel. For the latter, we characterize the underlying RKHS and conclude with a statistical error bound on its predictions.

\subsection{Gaussian Process Regression}
A \emph{Gaussian process} (GP) is a parameter-free regression model that allows the user to learn input-output mappings of the form $\hat f_d:\X\rightarrow\R$, $d\in\{1,\ldots,n\}$, from empirical data $\data=\{x_i,y_i\}_{i=1}^N$, $y_i:=[y_{i,1},\ldots,y_{i,n}]\T$, such that the prediction errors $\sum_{i=1}^N||y_{i,d}-\hat f_{d}(x_{i})||$ are minimized for all dimensions $d$ \citep{Rasmussen2018GP}.
Recall that, as specified in \eqref{eq:system}, we assume that the process noise on the samples $y_i$ is dimension-wise independent, which allows us to use individual GPs for each output dimension.
This is without loss of generality, and the results of this paper are applicable to non-diagonal covariance matrices by applying a linear transformation on the data gathered from the system.
In the following, we may drop the subscript $d$ indicating the dimension for simplicity.

As an infinite-dimensional generalization of the multivariate normal distribution, GPs are particularly popular for their ability to capture statistical information about the latent mapping. The following definition is analogous to Definition~2.2 by \cite{Kanagawa2018GPvsKernel}.
\begin{definition}[Gaussian process (GP)]\label{def:gp}
    Let $\hat k:\X\times\X\rightarrow\R$ be a (symmetric, positive-definite) \emph{kernel} or \emph{covariance function} and $\hat m_d:\X\rightarrow\R$.
     A random function $\hat f_d:\X\rightarrow\R$ is a \emph{GP}, denoted by $\hat f_d\sim\mathcal{GP}(\hat m_d,\hat k)$, if for any finite $X_N:=[x_1,\ldots,x_N]$, $x_i\in\X$, $N\in\N$, the vector $\hat f_{d,N}:=[\hat f_d(x_i)]_{i=1}^N$ has the distribution $$\textstyle\hat f_{d,N}\sim\mathcal{N}(\cdotx|\hat m_{d, N},\hat K_{N}),$$ with $\hat m_{d, N}:=[\hat m_d(x_i)]_{i=1}^N$ and Gram matrix $\hat K_{N}:=[\hat k(x_i,x_j)]_{i,j=1}^N$.
\end{definition}
The goal of GP regression is to generate predictions of the output at test points $x\in\X$ that are not captured by the data $\data$.
Even though we do not assume that the unknown data-generating function $f$ is a sample from the GP with kernel $\hat k$ (see Assumption~\ref{asm:knownkernel}), we can construct a so-called \emph{posterior GP} and obtain a probabilistic bound on its prediction error (e.g., via Theorem~2 by \cite{chowdhury2017kernelized} or the results by \cite{Fiedler2021GP}).
\begin{definition}[Posterior GP]\label{def:postGP}
    Let $\data=(X_N,Y_N)$ from the system \eqref{eq:system}. 
    Then, the \emph{posterior} distribution of $\hat f_d(x)$ given the data $\data$ is $\hat f_d(x) \sim\mathcal{N}(\cdotx| \hat\mu_{d,N}(x),\hat \sigma^2_{N}(x))$, characterized by the posterior mean and covariance functions
    \begin{align}
        \textstyle\hat \mu_{d,N}(x) &\textstyle:= \hat  k_{N}(x)\T [\hat K_{N} + \sigma_v^2 I_N]^{-1} Y_{d,N},\text{ and}\label{eq:posteriormean}\\
        \textstyle\hat \sigma^2_{N}(x) &\textstyle:= \hat k(x,x) - \hat k_{N}(x)\T [\hat K_{N} + \sigma_v^2 I_N]^{-1} \hat k_{N}(x),
        \label{eq:posteriorcov}
    \end{align}
    with $\hat k_{N}(x)\!:=\![\hat k(x_i, x)]_{i=1}^N$ and $Y_{d,N}$ the $d^{\text{th}}$ column of $Y_{N}$. 
\end{definition}
Based on Definition~\ref{def:postGP}, predictions of the next state of the system~\eqref{eq:system} for the current state $x\in\X$ can be generated via $\hat y_d(x)\sim\mathcal{N}(\cdotx|\hat\mu_{d,N}(x),\hat\sigma^2_{N}(x)+\sigma^2_{v})$.
We establish a probabilistic bound on the prediction error in Subsec.~\ref{sec:error} after introducing a specific type of GP in the upcoming subsection.


\subsection{Binary-Tree Gaussian Processes}\label{sec:btgp}
\vspace{-0.1cm}
We focus on GPs based on the \emph{binary-tree} (BT) kernel, which is defined on a finite space of binary strings and wields a range of practical advantages that renders it particularly interesting for formal approaches \citep{cohen2022log}.
Let us define a map $\mapq:\X\rightarrow\B^q$ encoding every state $x\in\X$ from the continuous space $\X$ as a binary string $s\in\B^q$ in the finite space of binary strings of length $q$.
Note that this implies a partitioning of the space $\X$ into $2^q$ disjoint sets, thus $q\in\N$ is called the \emph{bit-depth} or \emph{precision}.

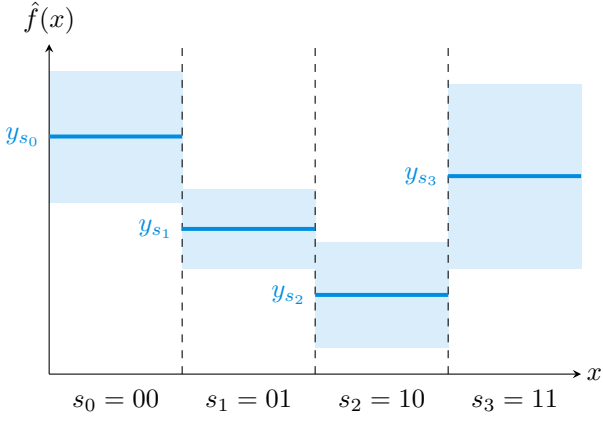
\begin{figure}
    \centering
    \begin{tikzpicture}[scale=.7]
        \tikzstyle{every node}=[font=\normalsize]
        \fill [fill=accentwashed] (10,13) rectangle (12.5,11.5);
        \fill [fill=accentwashed] (12.5,12) rectangle (15,10);
        \fill [fill=accentwashed] (15,15) rectangle (17.5,11.5);
        \fill [fill=accentwashed] (7.5,15.25) rectangle (10,12.75);
        \draw [arrows=-stealth] (7.5,9.5) .. controls (12.5,9.5) and (12.5,9.5) .. (17.5,9.5);
        \draw [arrows=-stealth] (7.5,9.5) .. controls (7.5,12.5) and (7.5,12.5) .. (7.5,15.75);
        \draw [dashed] (12.5,9.5) .. controls (12.5,12.5) and (12.5,12.5) .. (12.5,15.75);
        \draw [dashed] (10,9.5) .. controls (10,12.5) and (10,12.5) .. (10,15.75);
        \draw [dashed] (15,9.5) .. controls (15,12.5) and (15,12.5) .. (15,15.75);
        \draw [color=accent, line width=1.5pt] (7.5,14) .. controls (8.75,14) and (8.75,14) .. (10,14);
        \draw [color=accent, line width=1.5pt] (10,12.25) .. controls (11.25,12.25) and (11.25,12.25) .. (12.5,12.25);
        \draw [color=accent, line width=1.5pt] (12.5,11) .. controls (13.75,11) and (13.75,11) .. (15,11);
        \draw [color=accent, line width=1.5pt] (15,13.25) .. controls (16.25,13.25) and (16.25,13.25) .. (17.5,13.25);
        \node [font=\normalsize] at (8.75,9) {$s_0=00$};
        \node [font=\normalsize] at (11.25,9) {$s_1=01$};
        \node [font=\normalsize] at (13.75,9) {$s_2=10$};
        \node [font=\normalsize] at (16.25,9) {$s_3=11$};
        \node [font=\normalsize, color=accent, right] at (6.5,14) {$y_{s_0}$};
        \node [font=\normalsize, color=accent, right] at (9,12.25) {$y_{s_1}$};
        \node [font=\normalsize, color=accent, right] at (11.5,11) {$y_{s_2}$};
        \node [font=\normalsize, color=accent, right] at (14,13.25) {$y_{s_3}$};
        \node [font=\normalsize] at (17.75,9.5) {$x$};
        \node [font=\normalsize] at (7.5,16.25) {$\hat f(x)$};
    \end{tikzpicture}
    \caption{The BTGP contains functions that are piecewise constant over the partitions of $\X$. The posterior mean $\hat\mu_N$ and double standard deviation $\hat\mu_N\pm 2\hat\sigma_N$ are displayed as a piecewise-constant graph and an area plot in light blue, respectively. 
    For this example with chosen precision $q=2$, the posterior mean takes values $\hat\mu_N\in\{y_{s_0},\ldots,y_{s_3}\}$.
    }
    \label{fig:BTGP_fcns}
\end{figure}

In the following, we denote the set of states represented by a string $s\in\B^q$ as $\Xi_q(s)\subset\X$, i.e., we have $\mapq(x)=s$ for all $x\in\Xi_q(s)$.
It follows that the state space is the union of the partitions, i.e., $\X=\bigcup_{s\in\B^q}\Xi_q(s)$ for any precision $q\in\N$.
For a set $A\subset\X$, we may write $\mapq(A)\subset\B^q$ to denote its projection onto $\B^q$.
To simplify the notation, we may write $\map_i(x):=[\mapq(x)]_{\leq i}$, $x\in\X$, $i\in\N$, i.e.,
the operator that takes the first $i$ elements of the binary string $\mapq(x)$.
With this, we define the BT kernel formally.
\begin{definition}[Binary-tree (BT) kernel]\label{def:btkernel}
    Given a map $\mapq:\X\rightarrow\B^q$, 
    the \emph{BT kernel} $ \hat{k}_q:\X\times\X\rightarrow\R$ of \emph{bit-depth} or \emph{precision} $q\in\N$ is defined as
    \begin{equation}\textstyle
    \hat{k}_q(x,x') := \sum_{i=1}^q w_i\1\big(\map_i(x) = \map_i(x')\big),\label{eq:btkernel}
\end{equation}
with weight coefficients $w_i\in\R$ such that $\sum_{i=1}^q w_i=1$.
\end{definition}
Intuitively, the BT kernel $\hat{k}_q$ assigns its covariance mass based on whether points $x,x'\in\X$ fall into the same partitions. The final covariance value is a weighted sum of this coincidence on all discretization levels up to precision $q$.
It stands out that, in contrast to well-known alternatives such as the \emph{squared-exponential} or \emph{Mat\'{e}rn} kernel, the BT kernel is discontinuous since it is 
translation-\emph{variant}.
Hence, it produces discontinuous functional mappings when used in the context of GP regression. Fig.~\ref{fig:BTGP_fcns} depicts the piecewise-constant posterior mean $\hat \mu_N$ and double standard deviation $\hat \mu_N\pm 2\hat \sigma_N$ as functions of state of a GP utilizing a BT kernel of precision $q=2$. Moving forward, we will call such a GP a \emph{binary-tree Gaussian process} (BTGP), which yields inherently discrete abstractions/representations of the underlying system. 
Furthermore, we may decide to write them as explicit functions of $s\in\B^q$, i.e., $\hat \mu_N(s)$ and $\hat \sigma_N^2(s)$.
We note that although function samples from the BTGP (or any GP in general) do not live in the RKHS of the associated kernel almost surely, the resulting posterior mean function does \citep[Sec.~4]{Kanagawa2018GPvsKernel}.

\begin{remark}
 The BT kernel violates the conditions for a continuous RKHS \citep[Theorem~17, condition a)]{Berlinet2011RKHS}. Despite being discontinuous,
 the BT kernel is still a valid positive-definite kernel as shown in Proposition~1 by \cite{cohen2022log} and therefore by the Moore-Aronszajn theorem acts as a reproducing kernel for a unique RKHS \citep[Theorem~3]{Berlinet2011RKHS}, which contains only discontinuous functions.
\end{remark}


\begin{remark}
    An implication of using the BT kernel of precision $q$ in GP regression is that the resulting Gram matrix $\hat K_N$
    will be capped at a size of at most $2^q\times2^q$. This is since $M\in\N$ data samples $\{x_i,y_i\}_{i=1}^N$
    can be replaced by a modified dataset $\{x_s,y_s\}_{s\in\B^q}$, where $x_s\in \Xi_q(s)$ is a representative state and $y_s = \frac{1}{|A(s)|}\sum_{i\in A(s)} y_i$ with $A(s):=\{i\,|\,x_i\in\Xi_q(s)\}$ (averaging over $y_i$'s with $x_i$ being in the same partition set).
\end{remark}

The following theorem establishes a closed-form representation of the functions in the RKHS of the BT kernel.
\begin{theorem}[BT function space]\label{thm:fcns}
    Let $ \hat{k}_q:\X\times\X\rightarrow\R$ be the BT kernel in \eqref{eq:btkernel}.
    Then, its 
    associated RKHS
    contains only functions $f\in\Hilbert_{ \hat{k}_q}$ that can be represented as
    \begin{align}
        \textstyle f(x) \textstyle= \sum_{i=1}^q \sqrt{w_i} \sum_{s\in\B^i} y_{s} \1\big(x\in\Xi_i(s) \big),
        \label{eq:functionalrep}
    \end{align}
    for some $y_s\in\R$.
    The resulting finite dimensional RKHS $\Hilbert_{\hat{k}_q}$ (of dimension $\sum_{i=1}^q2^i$) is 
    characterized by the piecewise-constant feature map
    \begin{equation}
        \hat\phi_q(x) :=[\sqrt{w_i}\1(
        x\in\Xi_i(s)
        ) ]_{s\in\B^i, i\in\{1,2,\ldots,q\}}.
        \label{eq:feature_map}
    \end{equation}
\end{theorem}

A measure for the complexity of the functions in a given RKHS is the associated RKHS norm. For the BT kernel, a bound on the RKHS norm can be computed as follows.
\begin{corollary}[BT RKHS norm bound]\label{thm:rkhsnorm}
    Let $\hat{k}_q:\X\times\X\rightarrow\R$ be the BT kernel in \eqref{eq:btkernel} and $\Hilbert_{\hat{k}_q}$ its associated RKHS.
    For all functions $f\in\Hilbert_{\hat{k}_q}$ of the form \eqref{eq:functionalrep},
    we have that
    \begin{equation*}
        \textstyle||f||_{\Hilbert_{\hat{k}_q}} \leq \sqrt{\sum_{i=1}^q w_i \sum_{s\in\mathbb{B}^i} y_{s}^2} .\label{eq:rkhsnorm}
    \end{equation*}
\end{corollary}

\subsection{Statistical Error Bound}\label{sec:error}
\vspace{-0.1cm}


To use GP regression in the context of formal approaches, we must quantify the error between the unknown data-generating function $f\in{\Hilbert_{ k}}$ in \eqref{eq:system} and the GP posterior mean $\hat\mu_N\in{\Hilbert_{\hat{k}_q}}$ in \eqref{eq:posteriormean}.
To be able to address a wealth of functions $f$ beyond those discontinuous functions living in the RKHS ${\Hilbert_{\hat{k}_q}}$ associated with the BT kernel, we leverage Assumption~\ref{asm:knownkernel} to quantify the additional error introduced by the erroneous function space.
The following theorem takes inspiration from Theorem~2.1 by \cite{hsu2012tail} and Proposition~2 by \cite{Fiedler2021GP}, where latter is itself an adaptation of Theorem~2 by \cite{chowdhury2017kernelized}.
As mentioned before, the output dimensions -- indexed by $d$ -- are modeled by one GP each.
\begin{theorem}[Piecewise-constant error bound]\label{thm:error}
    \hfill Consider\break the BT kernel $\hat{k}_q$ in \eqref{eq:btkernel} with precision $q\in\N$ and the associated RKHS $\Hilbert_{\hat{k}_q}$.
    Let the unknown function $f\in\Hilbert_{ k}$ with a known translation-invariant kernel $k$ that has a constant $c$ satisfying $c^2\geq k(x,x)\geq 0$ for all 
 $x\in\X$, and a complexity bound $B\geq||f||_{\Hilbert_{ k}}\geq0$ (Assumption~\ref{asm:knownkernel}).
    Given data $\data=(X_N,Y_N)$ and a constant $\delta\in(0,1)$, we have for each dimension $d\in\{1,\ldots,n\}$ for the corresponding BTGP posterior mean $\hat\mu_{N,d}(s)\in\Hilbert_{\hat{k}_q}$ in \eqref{eq:posteriormean} that
    \begin{equation}
        \P\Big( \forall s\in\B^q:\, \sup_{x\in\Xi(s)} \big|\hat\mu_{N,d}(s)-f_d(x)\big| \leq \varepsilon_d(s) \Big) \geq 1-\delta,\label{eq:learning_error}
    \end{equation}
    with the error $\varepsilon_d(s):=\varepsilon_{d,1}(s)+\varepsilon_{d,2}(s)+\varepsilon_{d,3}(s)$ given by
    \begin{align*}
    \textstyle\varepsilon_{d,1}(s) &:= \min\Big\lbrace\textstyle\big|\big|\hat C_N^{-1}\hat k_{N}(s)\big|\big| \sqrt{N+2\left(N\log\frac{1}{\delta}\right)^{\frac{1}{2}}+2\log\frac{1}{\delta}}, \Big.\\
    & \textstyle\Big. \frac{\hat\sigma_{N,d}(s)}{\sigma_v} \textstyle
        \sqrt{\trace{\Sigma}\!+\!2\left(\trace{\Sigma^2}\log\frac{1}{\delta}\right)^{\frac{1}{2}}\!+\!2||\Sigma||\log\frac{1}{\delta}}
        \Big\rbrace,\\
    \textstyle\varepsilon_{d,2}(s)  &:=  \textstyle B\big(2\left| c^2 - k^{+}_{ x_s}(s) \right|\big)^{\frac{1}{2}},\\
    \textstyle\varepsilon_{d,3}(s) &:= \textstyle B\left|\hat k_N(s)\!\T\!\hat C_N^{-1}\!\!\left(\!K_N \hat C_N^{-1}\hat k_N(s)\!-\!2k_N(x_s)\!\right)\!\!+\!c^2\right|^{\frac{1}{2}},
    \end{align*}
    where $\hat C_N:=[\hat K_{N}+\sigma_v^2 I_N]$, $\hat k_N(s):=[k(x_i,\mapq(s))]_{i=1}^N$, the posterior covariance $\hat\sigma^2_{N,d}(s)$ in \eqref{eq:posteriorcov}, $\Sigma := \hat K_N\hat C_N^{-1}$, $k^{+}_{x_s}(s):=\sup_{x\in\Xi(s)}k(x_s,x)$ for some fixed set of representative points $\{ x_s \}_{s\in\B^q}\subset \X$, $K_{N}:= [k(x_i,x_j)]_{i,j=1}^N$, and $k_{N}(x):= [k(x_i,x)]_{i=1}^N$.  
\end{theorem}
\begin{remark}[Approximating a continuous kernel]\label{rem:approx_continuous_kernel}
    The performance benefits of GP regression via the BT kernel stem from the fact that it constitutes a series of subsequent partitionings of the input space such that in every discretization level the precision of the previous partitioning is refined (or kept the same). For each level, functional mass represented by the corresponding weight coefficient is applied whenever two points fall into the same partition. As shown in Theorem~1 by \cite{cohen2022log}, this allows for an efficient computation of the Gram matrix and its inverse (computations for the inversion have a complexity of $O(N)$ for every precision level $i:=1,\ldots,q$ instead of $O(iN)$, which overall leads to $O(qN)$).
    
\begin{remark}\label{rem:gram_matrix_size}
    The theorem above establishes a finite-sample bound on the error between the posterior mean of the BTGP and the actual unknown dynamics $f$. The complexity bound $B$ on $||f||_{\Hilbert_{ k}}$ enables us to provide guarantees even when $f$ does not live in the RKHS of the BTGP, $\Hilbert_{\hat{k}_q}$, e.g., if $f$ is continuous. This is called the `misspecified case'.
    In comparison to \cite{Fiedler2021GP}, we quantify the introduced error $\varepsilon_d$ without computing a global upper bound on the error between the two kernels, which would be at its maximum in our case as $\hat k_q$ does generally not approximate $k$. 
%
The first term in the minimization of $\varepsilon_{d,1}$ is equivalent to prior works and captures the statistical error of the properly specified case, which is generally very conservative. We provide an alternative second term in an effort to reduce this error by exploiting the low-dimensional BT kernel Gram matrices.
Nevertheless, note that whilst the error bounds $(\varepsilon_{2,d}, \varepsilon_{3,d})$ can be tightened by increasing the precision $q$, this is not the case for $\varepsilon_{d,1}$.
\end{remark}


Determining the error bound $\varepsilon_{d,3}$ requires us to compute Gram matrices via both $k$ and $\hat{k}_q$, 
reducing the scalability benefits of the BTGP. 
It is interesting to study the quantification of $\varepsilon_d$ without the need for constructing $K_N$.

    One might be tempted to use the BT kernel in Definition~\ref{def:btkernel}, or an analogously defined alternative kernel bearing the same properties, to approximate a continuous kernel such as the squared-exponential kernel for the limit case $q\rightarrow\infty$, allowing for a simplification of Theorem~\ref{thm:error}. However, such a kernel does not exist. In fact, let $\hat{k}_q(x,x')$ be such a kernel with precision $q\in\N$ and consider $x\in\X$ to be fixed and $x'\in\X$ variable. Then, for $q<q'\in\N$ the kernels $\hat{k}_{q}$ and $\hat{k}_{q'}$ define two discrete distance metrics where the exponential increase in the level of precision of $\hat{k}_{q'}$ compared to $\hat{k}_{q}$ is focused solely on the neighborhood of $x$. 
    Hence, escalating the level of precision merely increases the smoothness of the covariance in the neighborhood of $x$ and has no effect on the quantification elsewhere. 
\end{remark}

In contrast to methods such as probabilistic coupling relations \citep{schon2023bayesian} where the error between the true system and the discrete abstraction is captured by computing local parameters bounding the local deviations, the error bound $\varepsilon_d(s)$ in Theorem~\ref{thm:error} summarizes the error in a formula that similarly yields local deviations when evaluated for individual partitions $s\in\B^q$.




\section{Verification Approach}\label{sec:verif}
\vspace{-0.2cm}
The verification approach we consider is based on the construction of a finite-state abstraction of the dynamical system in \eqref{eq:system} via the BTGP introduced in Sec.~\ref{sec:btgp}. 
For the learned BTGPs, let $\B^q$ be a finite set of binary strings representing the states of this abstraction.
Recall that a BTGP defines a map $\mapq:\X\rightarrow\B^q$
that maps every state $x\in\X$ from the continuous state space $\X$ to a string $s\in\B^q$ associated with a partition of $\X$ and a representative state $x_s\in\X$. 
To capture the probabilistic ambiguity w.r.t. transitions between discrete partitions, recent approaches for verifying infinite-horizon specifications are based on \emph{interval Markov chains} (IMCs), which can be learned from data via GP regression \citep{Badings2022epistemicUncert, Lavaei2022DDMDP,Jackson2020safety}. 
In this manuscript, we follow a similar approach,
however, we perform GP regression using the BT kernel introduced in Sec.~\ref{sec:btgp}, that allows us to reduce the computational effort in constructing an IMC from data significantly.

\begin{definition}[Interval Markov chain (IMC)]
    An IMC is a tuple $\widehat\M:=(\S,s_{\text{init}},
    \underline\Tr,\bar\Tr)$, with
    a finite state space $\S$ of
    discrete states $s\in\S$;
    initial state $s_{\text{init}}\in\S$;
    and probability bounds $\underline\Tr,\bar\Tr:\S\times\S\rightarrow[0,1]$
    describing lower and upper bounds on the transition probability.
\end{definition}

\begin{figure}
\centering
    \includegraphics[width=\columnwidth]{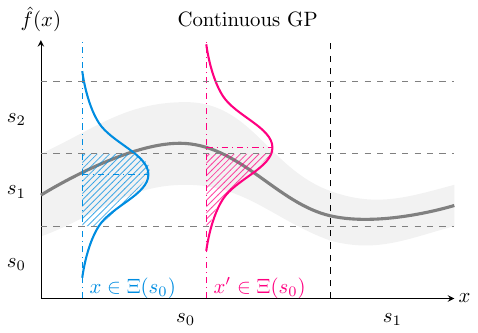}
    \includegraphics[width=\columnwidth]{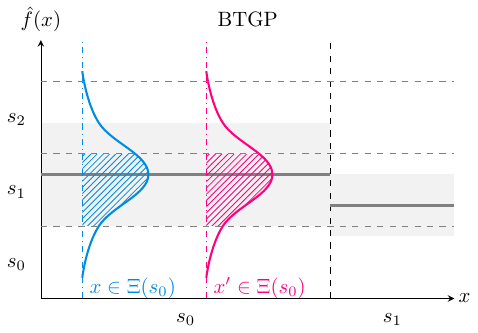}


\caption{Computing transition probabilities from a continuous GP (top) vs. from a BTGP (bottom).
}
\label{fig:int_GP_BTGP}
\end{figure}

\subsection{Computing the Abstract IMC via BTGP}
\label{sec:trans_via_btgp}
The abstract IMC $\widehat\M:=(\S,s_{\text{init}},
    \underline\Tr,\bar\Tr)$
    is constructed with the state space $\S:=\B^q$, and the initial state $s_{\text{init}}:=\mapq(x_{\text{init}})$ with $x_{\text{init}}$ being the initial state of the system $\M$.
Recall that we use one BTGP to model the dynamics for each output dimension. We hence obtain $n$ BTGPs $\hat f_d\sim\mathcal{GP}(\hat\mu_{N,d},\hat\sigma_{N,d}^{2})$, $d\in\{1,\ldots,n\}$, where we use the same precision $q\in\N$ to generate a common partitioning $\S=\mapq(\X)$.
Computing the discrete transition kernels $\underline\Tr,\bar\Tr$ from the constructed BTGPs is comparatively straightforward due to the piecewise-constant nature of the posterior mean and covariance functions of the BTGP. In other words, for every partition $s\in\S$, the consecutive state (for each dimension $d\in\{1,\ldots,n\}$) is distributed according to a Gaussian distribution whose mean is uncertain in the form of $\hat\mu_{N,d}(s)+\epsilon_d(s)$ with error $\epsilon_d(s)\in E_d(s):=[-\varepsilon_d(s),\varepsilon_d(s)]$ obtained via the corresponding stochastic learning errors (Theorem~\ref{thm:error}). The Gaussian distribution has the covariance $\hat\sigma^{2}_{N,d}(s)$ (see Fig.~\ref{fig:int_GP_BTGP}, bottom). 
We obtain lower and upper bounds on the probability of transitioning from $s$ to any $s'\in\S$ as
\begin{align}
    & \underline\Tr(s'| s) \!=\!\!\! \min_{\epsilon\in E(s)} g(s,\epsilon) \,\,\text{ and }\,\, \bar\Tr(s'| s) \!=\!\!\! \max_{\epsilon\in E(s)} g(s,\epsilon),\label{eq:ublb_trans_prob_btgp}\\
    & g(s,\epsilon) := \int_{x'\in\Xi(s')}\! \prod_{d=1}^n\!\mathcal{N}(dx'_d|\hat\mu_{N,d}(s)\!+\!\epsilon_d,\!\hat\sigma^{2}_{N,d}(s)),\label{eq:function_g}
\end{align}
with $E(s) := \prod_{d=1}^n E_d(s)$, $\epsilon = [\epsilon_d]_{d=1}^n$, and $x' =[x'_d]_{d=1}^n$.
If the partitions of the BTGPs are selected to be rectangular, the integration in \eqref{eq:function_g} and subsequent optimizations in \eqref{eq:ublb_trans_prob_btgp} can be done separately for each dimension.
In comparison, for a kernel that is not piecewise-constant, e.g., the squared-exponential kernel in Fig.~\ref{fig:int_GP_BTGP} (top), obtaining the bounds requires optimizing with respect to $\epsilon$ the function
\begin{equation}
    g(s,\epsilon)\!=\! \underset{x\in\Xi(s)}{\optimize}\!\!\int_{x'\in\Xi(s')}\! \prod_{d=1}^n\!\mathcal{N}(dx'_d|\hat\mu_{N,d}(x)+\epsilon_d(x),\!\hat\sigma^{2}_{N,d}(x)),\label{eq:integration_gp}
\end{equation}
with $\optimize\in\{\max,\min\}$ for each pair $(s,s')$.
%
This shows that the BTGP eliminates the additional inner optimization, thus significantly simplifying the computational burden associated with constructing the IMC to one integration per bound per transition. 



\subsection{Interval Iteration Algorithm for Verification}
Once the abstract IMC is constructed via \eqref{eq:ublb_trans_prob_btgp}-\eqref{eq:function_g}, the \emph{interval iteration algorithm} from the work by \cite{haddad2018interval} can be employed to compute lower and upper bounds on the reachability probability. The algorithm
is based on the iterative update of two \emph{value functions} forming an envelope or \emph{interval} around the true probability of satisfying the specification starting from some initial partition.
\iflong We defer the details to Appendix~\ref{app:algorithm}.
\fi


With the next theorem, we provide a robust satisfaction probability for the latent true system in \eqref{eq:system} satisfying a reachability specification $\psi_\target$ as follows.
\begin{theorem}[Robust satisfaction probability]\label{thm:satProb}
    Consider the setup of Theorem~\ref{thm:error}.
    Let $\hat f_d \sim\mathcal{GP}(\hat\mu_{N,d},\hat\sigma^2_{N,d})$, $d\in\{1,\ldots,n\}$ be the posterior BTGPs based on data $\data$ from the system $\M$ in \eqref{eq:system}.
    Let $\widehat\M$ be the corresponding IMC constructed as outlined in \eqref{eq:ublb_trans_prob_btgp}.
    Then, we have for an infinite-horizon reachability specification $\psi_\target$ and $\M$ initialized at any state $x_{\text{init}}\in\X$ that
    \begin{equation*}
        V_{\text{min}}(s_{\text{init}}) \leq \P(\M\vDash\psi_\target) \leq V_{\text{max}}(s_{\text{init}}),
    \end{equation*}
    where $s_{\text{init}}:=\mapq(x_{\text{init}})$ and $V_{\text{min}}$, $V_{\text{max}}$ are the lower and upper bounds on the value function obtained via the interval iteration algorithm  
    \iflong (Algorithm~\ref{alg:safety}).
    \else \citep{haddad2018interval}.
    \fi
\end{theorem}

\section{Case Study}
\label{sec:casestudies}
\begin{figure}
    \centering
    \includegraphics[width=\columnwidth]{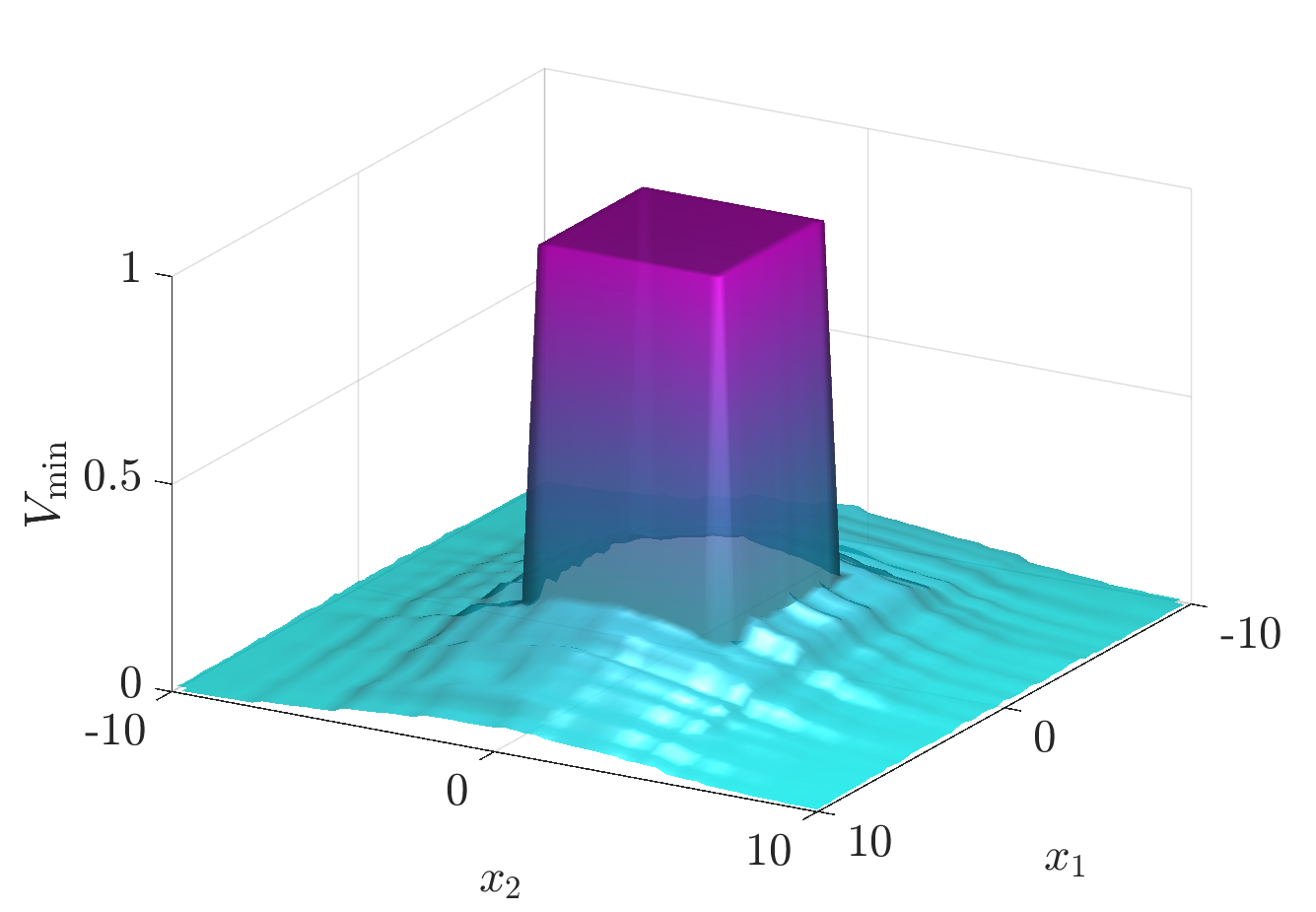}
    \vspace{-0.1cm}
    \caption{Lower bound on the infinite-horizon reachability probability as a function of the initial state. 
    }
    \vspace{-0.1cm}
    \label{fig:satProb_casestudy}
\end{figure}

\vspace{-0.1cm}
Consider the discrete-time nonlinear system
\begin{equation*}
    \begin{bmatrix}
        x_{1, t+1}\\
        x_{2, t+1} 
    \end{bmatrix} = \begin{bmatrix}
        x_{1,t} - \tau_s x_{1,t} + 0.5\tau_s \sin(x_{2,t})\\
        x_{2,t} - \tau_s x_{2,t} + 0.5\tau_s \sin(x_{1,t})
    \end{bmatrix} + v_t,
\end{equation*}
evolving on the continuous state space $\X=[-10,10]^2\subset\R^2$ with $\tau_s=0.5$ and $v_t\sim \mathcal{N}(\cdotx|0,\sigma_v^2 I_2)$.
The goal is to compute a lower bound on the probability of the system reaching the target set $\target=[-3, 3]^2\subset\X$ based on $N=5000$ observations $\data$ from the system generated with $\sigma_v=3.16$.
We construct two BTGPs -- one for each output dimension -- with a precision of $q=12$.
Training and evaluation of the BTGPs at the representative states takes $13$ seconds on an Apple MacBook Pro M1.
To compute an error bound via Theorem~\ref{thm:error}, we choose 
the confidence bound $\delta:=0.2$, squared-exponential kernels
$k_d(x,x'):=c_d^2\exp(-\frac{1}{2}(x-x')\T M_d^{-1}(x-x'))$ for $d\in\{1,2\}$ with hyperparameters $c_1=12$, $c_2=7$, and $M_d=\mathrm{diag}(l_d^2)$ where $l_1=[4000, 2500]$ and $l_2=[500, 2000]$,
and dimension-wise complexity bounds $B_1=0.015$, $B_2=0.006$.
For $\{ x_s\}_{s\in\B^q}$, we choose the center points of the partitions.
Due to its piecewise-constant nature, generating the IMC via \eqref{eq:ublb_trans_prob_btgp} from the BTGPs and error bounds takes less than $3$ seconds.
In comparison, abstraction of a comparable continuous GP via \eqref{eq:integration_gp} takes around $8$ hours.
The IMC is subsequently verified via the interval iteration algorithm, which converges after $9$ seconds 
    \iflong ($\nu=10^{-8}$). 
    \fi
The resulting lower bound on the probability of reaching $\target$ is depicted as a function of the initial state in Fig.~\ref{fig:satProb_casestudy}.


\section{Conclusion}\label{sec:conclusion}
In this paper, we showcased how to use binary-tree Gaussian processes (BTGPs) for learning model representations from data that are naturally ideal for finite-state abstraction and verification due to the piecewise-constant nature of the employed binary tree kernel. We also formulated the error bounds in the learning, which were used to generate the abstraction as an interval Markov chain.
We are currently working on improving the computation of the error bounds and benchmarking the proposed approach against alternative data-driven methods.





\bibliography{root}

\iflong 
\appendix
\section{Additional Figures on the Case Study}
Fig.~\ref{fig:btgp_casestudy} illustrates the fit of the BTGPs for the system in the case study (Sec.~\ref{sec:casestudies}). The top and bottom plots show, respectively for dimensions $d=1,2$, the posterior mean and triple standard deviation of the BTGPs as well as the actual model response for reference.
\begin{figure}
    \centering
    \includegraphics[width=\columnwidth]{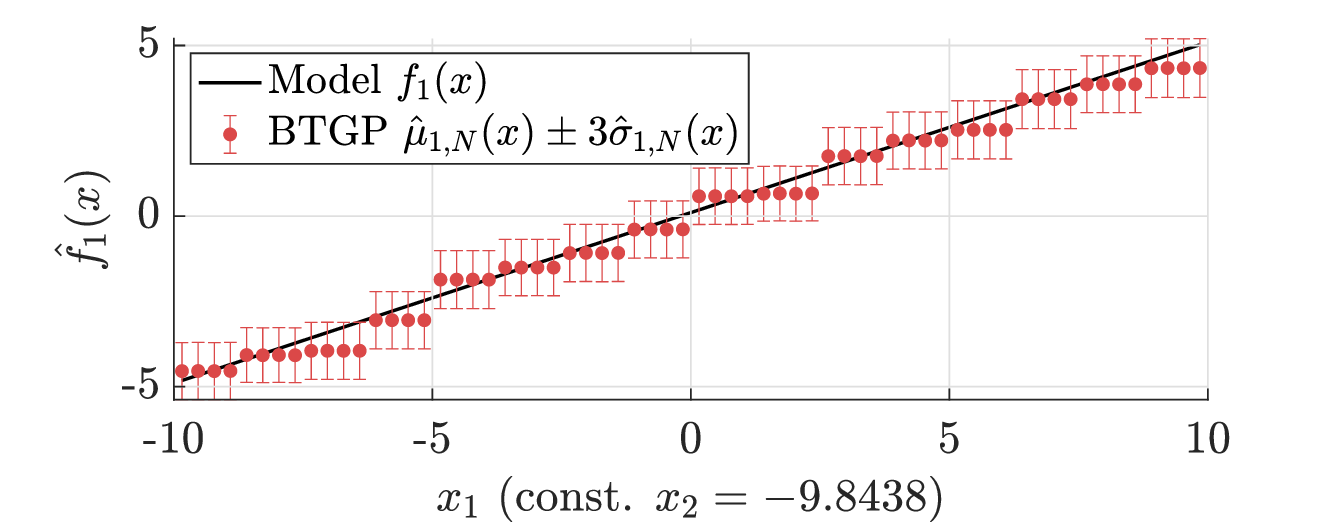}
    \includegraphics[width=\columnwidth]{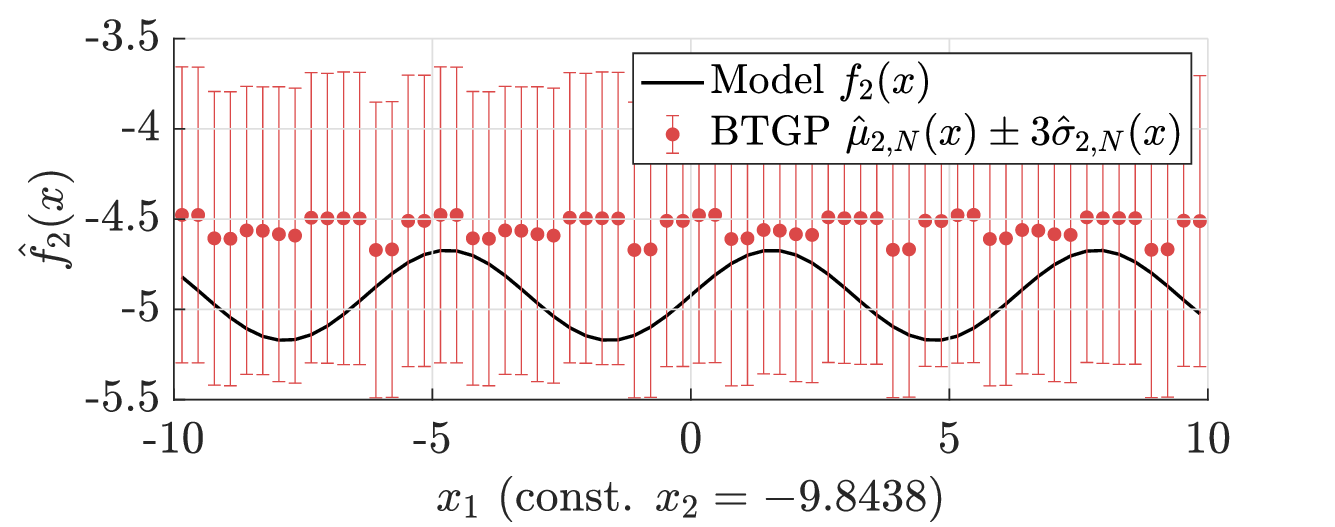}
    \caption{BTGP model fit $\hat \mu_{d, N}(x)\pm 3\hat\sigma_{d, N}(x)$ versus $f_d(x)$ (top: $d=1$, bottom: $d=2$).
    }
    \label{fig:btgp_casestudy}
\end{figure}

\section{Proof of Theorem~\ref{thm:fcns}}
We can write the BT kernel in Eq.~\eqref{eq:btkernel} as the sum
\begin{equation}
    \hat{k}_q(x,x') = \sum_{i=1}^q \kappa_i(x,x'),\label{eq:weightedsum}
\end{equation}
of kernels $\kappa_i(x,x') := \phi_i(x)\T\phi_i(x)$ with feature maps
\begin{equation*}
    \phi_i(x):=\sqrt{w_i}\left[\1(
    x\in\Xi_i(s)
    )\right]_{s\in\B^i},
\end{equation*}
where we expanded each $i^\text{th}$ summand of $\hat{k}_q$ w.r.t. all possible binary substrings of length $i$.
Note that the binary substrings $s\in\B^i$ in $\kappa_i:\X\times\X\rightarrow[0,1]$ are unique and $\kappa_i$ is clearly positive-definite.
The associated RKHS $\Hilbert_{\kappa_i}$ is of dimension $2^i$.
Since finite dimensional spaces are automatically complete, $\Hilbert_{\kappa_i}$ is simply the span of the functions $\kappa_i(x,\cdotx)$, $\forall x\in\X$.
All functions $f_i\in\Hilbert_{\kappa_i}$ are hence linear combinations of the form
\begin{equation*}
    f_i(x) = \sum_{j=1}^m \alpha_j \kappa_i(x,x_j)
    = \sum_{j=1}^m \alpha_j \phi_i(x)\T\phi_i(x_j),
\end{equation*}
for arbitrary $m\in\N$, $\alpha_j\in\R$, and $x_j\in\X$ \citep{Berlinet2011RKHS}. Note that $\phi_i(x_j)$ only places functional weight on the partition $s=\map_i(x_j)$.
Therefore, $f_i\in\Hilbert_{\kappa_i}$ is always of the form 
\begin{equation*}
    f_i(x) 
     = y_i\T \phi_i(x),
\end{equation*}
with $y_i\in\R^{2^i}$. 
Since the BT kernel is a weighted sum of kernels $\kappa_i$ as shown in \eqref{eq:weightedsum}, we obtain 
\begin{equation*}
    f(x) = y\T \hat\phi_q(x),
\end{equation*}
with $y\T:=[y_1\T,\ldots,y_q\T]$ and feature map $\hat\phi_q(x)\T:=[\phi_1(x)\T,\ldots,\phi_q(x)\T]$ via the kernel sum rule \citep[Theorem~5]{Berlinet2011RKHS}, which is equivalent to \eqref{eq:functionalrep}.
Note that $\hat{k}_q(x,x')=\innerH{\hat\phi_q(x)}{\hat\phi_q(x)}{\Hilbert_k}$ and $\Hilbert_{\hat{k}_q}$ is $(\sum_{i=1}^q2^i)$-dimensional.

\section{Proof of Corollary~\ref{thm:rkhsnorm}}
We can write the BT kernel in \eqref{eq:btkernel} as the sum
\begin{equation*}
    \hat{k}_q(x,x') = \sum_{i=1}^q w_i \sum_{s\in\mathbb{B}^i} \kappa_s(x,x'),
\end{equation*}
of (positive-definite) kernels 
\begin{equation*}
    \kappa_s(x,x') := \1(\map_{|s|}(x) = s) \1(\map_{|s|}(x') = s).
\end{equation*}
Following similar reasoning as in the proof of Theorem~\ref{thm:fcns}, we find
\begin{align*}
    \Hilbert_{\kappa_s} &= \left\lbrace f_s\, |\, f_s = y_s \1(\map_{|s|}(\cdotx) = s), \, y_s\in\R \right\rbrace,\\
    &= \left\lbrace f_s\, |\, f_s = y_s \1(\map_{|s|}(\cdotx) = s)\1(s_{\leq |s|} = s), \, y_s\in\R \right\rbrace,\\
    &= \left\lbrace f_s\, |\, f_s = y_s \kappa_s(\cdotx,x_s), \, y_s\in\R,\, x_s\in\Xi_{|s|}(s) \right\rbrace.
\end{align*}
We compute the norm of $f_s\in\Hilbert_{\kappa_s}$ given $\kappa_s$ via
\begin{equation*}
    ||f_s||_{\Hilbert_{\kappa_s}}^2 \equiv \innerH{f_s}{f_s}{\Hilbert_{\kappa_s}}
    = y_s^2 \innerH{\kappa_s(\cdotx,x_s)}{\kappa_s(\cdotx,x_s)}{\Hilbert_{\kappa_s}} = y_s^2,
\end{equation*}
where we used $\kappa_s(x_s,x_s)=1$.
From Aronszajn's sums of kernels theorem \citep[Theorem~5]{Berlinet2011RKHS} we have $\forall f\in\Hilbert_{\hat{k}_q}$
\begin{equation*}
    ||f||_{\Hilbert_{\hat{k}_q}}^2 
    \leq \sum_{i=1}^q w_i \sum_{s\in\mathbb{B}^i} ||f_s||_{\Hilbert_{\kappa_s}}^2,
\end{equation*}
concluding the proof. 

\section{Proof of Theorem~\ref{thm:error}}
    Similar to the proof of Theorem~2 by \cite{chowdhury2017kernelized}, we start by splitting the error bound into one term related to the noise and one to the statistical approximation error. For the data $\data=(X_N,Y_N)$, let $Y_d:=[y_{1,d},\ldots,y_{N,d}]\T$ and recognize that we have $Y_d\equiv F_d+V_d$, with $F_d:=[f_d(x_1),\ldots,f_d(x_N)]\T$ and $V_d:=[v_{1,d},\ldots,v_{N,d}]\T$. Via Cauchy-Schwarz we obtain
    \begin{align}
    \begin{split}
        &\big|\hat\mu_{N,d}(s)-f_d(x)\big|\\
        &\hspace{15pt}\leq
        \big| \hat k_N(s)\T \!\hat C_N^{-1}V_{d} \big|
        + \big| \hat k_N(s)\T \!\hat C_N^{-1}F_{d}-f_d(x) \big|.
    \end{split}
        \label{eq:error_temp}
    \end{align}
    where we abbreviate $\hat C_N^{-1}:=[\hat K_N+\sigma_v^2 I_N]^{-1}$.
    We bound the noise term following similar steps to Proposition~2 by \cite{Fiedler2021GP} to get $$\textstyle\big| \hat k_N(s)\T\hat C_N^{-1}V_{d} \big| \leq \big|\big|\hat C_N^{-1}\hat k_{N}(s)\big|\big| \sqrt{N+2\left(N\log\frac{1}{\delta}\right)^{\frac{1}{2}}+2\log\frac{1}{\delta}},$$
    which yields the first term in the minimization of $\varepsilon_{d,1}(s)$.
    As an alternative bound, 
    we can follow the same steps as in the proof of Theorem~2 by \cite{chowdhury2017kernelized} to derive that
    $$|\hat k_N(s)\T \hat C_N^{-1}V_d|\leq \sigma_v^{-1}\hat\sigma_{N,d}(s)\sqrt{V_d\T\hat K_N \hat C_N^{-1}V_d}.$$
    From \citep[Theorem~2.1]{hsu2012tail} we have that
    $$\P\!\left(||\Sigma V_d||^2 \leq \textstyle
            \trace{\Sigma}\!+\!2\left(\trace{\Sigma^2}t\right)^{\frac{1}{2}}\!+\!2||\Sigma||t\right)\geq 1-e^{-t}\!,$$
    where $\Sigma := \hat K_N\hat C_N^{-1}$. Setting $t:=\log\left(\frac{1}{\delta}\right)$ yields the second term in the minimization of $\varepsilon_{d,1}(s)$.
    Depending on the setting, one or the other bound will provide a less conservative error for $\varepsilon_{d,1}(s)$.

    Next, we select some fixed $ x_{s}\in\X$ for all $s\in\S$ such that $\bar f_d(s):=f_d( x_{s})$.
    Expanding the second term in \eqref{eq:error_temp} with $\bar f_d(s)$ and Cauchy-Schwarz yields for every $s\in\S$ that for all $x\in\Xi(s)$
    \begin{align}
    \begin{split}
        &\big| \hat k_N(s)\T \!\hat C_N^{-1}F_{d}-f_d(x) \big| \\
        &\hspace{15pt}\leq \big| \hat k_N(s)\T \!\hat C_N^{-1}F_{d}-\bar f_d(s) \big|
        +\big| \bar f_d(s) - f_d(x) \big|.
    \end{split}\label{eq:error_temp2}
    \end{align}
    The latter term is bounded from above via
    \begin{align*}
        \big| \bar f_d(s) - f_d(x) \big| &= \big| (\phi( x_s)-\phi(x))\T f_d \big|,\\
        &\leq \big|\big| \phi( x_s)-\phi(x)\big|\big|_{\Hilbert_k} \big|\big| f_d\big|\big|_{\Hilbert_k},\\
        &=  \big|k( x_s,\bar x_s)\!-\!2k( x_s,x)\!+\!k(x,x)\big|^{\frac{1}{2}} \big|\big| f_d\big|\big|_{\Hilbert_k},\\
        &\leq \varepsilon_{d,2}(s).
    \end{align*}
    
    Let $\Phi_N:=[\phi(x_1),\ldots,\phi(x_N)]\T$.
    With this and Cauchy-Schwarz, the prior term in \eqref{eq:error_temp2} yields
    \begin{align*}
        &\big| \hat k_N(s)\T \hat C_N^{-1}F_{d}-\bar f_d(s) \big| 
        \\&\hspace{65pt}\leq \big|\big| \hat k_N(s)\T \hat C_N^{-1}\Phi_N -\phi( x_s)\T \big|\big|_{\Hilbert_k} \big|\big|f_d\big|\big|_{\Hilbert_k},
    \end{align*}
    which reduces to $\varepsilon_{d,3}(s)$, concluding the proof.



\section{Interval iteration algorithm}
\label{app:algorithm}
In this section, we provide the details of the interval iteration algorithm by \cite{haddad2018interval} (Algorithm~\ref{alg:safety}). Inter alia, they establish the following result.
\begin{proposition}[Value convergence]\label{prop:value}
    For a given IMC $\widehat\M:=(\S,s_{\text{init}},
    \underline\Tr,\bar\Tr)$ and an infinite-horizon reachability specification $\psi_\target$, let $V^*:\S\rightarrow[0,1]$ be the \emph{optimal value function} such that $\P(\widehat\M\vDash\psi_\target) = V^*(s_{\text{init}})$.
    Then, the interval iteration in Algorithm~\ref{alg:safety} converges to $V^*$ as $i\rightarrow\infty$.
    Furthermore, for every iteration $i\in\N\cup\{0\}$ we have $\underline V_i(s_{\text{init}})\leq V^*(s_{\text{init}})\leq \bar V_i(s_{\text{init}})$.
\end{proposition}
\begin{algorithm}
    \caption{Interval iteration algorithm 
    }
    \label{alg:safety}
    \begin{algorithmic}[1] 
    \Statex\rule[.5ex]{\linewidth}{1pt}
        \Statex\textbf{Input:} IMC $\widehat{\M}=(\S, s_{\text{init}}, \underline{\Tr}, \overline{\Tr})$, specification $\psi_{\target}$, stopping threshold $\nu>0$
        \Statex\rule[.5ex]{\linewidth}{1pt}
        \For{$s \in \S$} 
            \State $\overline{r}(s)\! \gets\! \min\!\left\lbrace \sum_{s'\in\mapq(\target)}\overline{\Tr}(s' \vert s),\, 1\right\rbrace$
            \State $\underline{r}(s)\! \gets\! \sum_{s'\in\mapq(\target)}\underline{\Tr}(s' \vert s)$
            \State $\overline{l}(s)\! \gets\! 1 -\sum_{s'\in\S}\underline{\Tr}(s' \vert s)$
            \State $\underline{l}(s)\! \gets\! \max\!\left\lbrace 0,\, 1- \sum_{s'\in\S}\overline{\Tr}(s' \vert s) \right\rbrace$
        \EndFor
        
        \State $i\gets0$, $\overline{V}_0 \gets 1$, $\underline{V}_0 \gets 0$
        
        \While{$\| \overline{V}_i - \underline{V}_i \|_\infty > \nu$}
            \State $r, \Tr \gets$ solve min LP in \eqref{eq:LP}
            \For{$s\in\S$}
                \State $\overline{V}_{i+1}(s) \gets r(s) + \sum_{s'\in\S\setminus\mapq(\target)}\Tr(s' \vert s)\overline{V}_i(s')$
                \State $\underline{V}_{i+1}(s) \gets r(s) + \sum_{s'\in\S\setminus\mapq(\target)}\Tr(s' \vert s)\underline{V}_i(s')$
            \EndFor
            \State $i\gets i+1$, $\overline{V}_i \gets \overline{V}_{i+1}$, $\underline{V}_i \gets \underline{V}_{i+1}$
        \EndWhile
        
        \State $V_{\text{min}} \gets \underline{V}_i$ \Comment{Take underapproximation}
        \State $i\gets0$, $\overline{V}_0 \gets 1$, $\underline{V}_0 \gets 0$
        
        \While{$\| \overline{V}_i - \underline{V}_i \|_\infty > \nu$}
            \State $r, \Tr \gets$ solve max LP in \eqref{eq:LP}
            \For{$s\in\S$}
                \State $\overline{V}_{i+1}(s) \gets r(s) + \sum_{s'\in\S\setminus\mapq(\target)}\Tr(s' \vert s)\overline{V}_i(s')$
                \State $\underline{V}_{i+1}(s) \gets r(s) + \sum_{s'\in\S\setminus\mapq(\target)}\Tr(s' \vert s)\underline{V}_i(s')$
            \EndFor
            \State $i\gets i+1$, $\overline{V}_i \gets \overline{V}_{i+1}$, $\underline{V}_i \gets \underline{V}_{i+1}$
        \EndWhile
        
        \State $V_{\text{max}} \gets \overline{V}_i$ \Comment{Take overapproximation}
        \Statex\rule[.5ex]{\linewidth}{1pt}
        \Statex \textbf{Output:} $V_{\text{min}}, V_{\text{max}}$
        \Statex\rule[.5ex]{\linewidth}{1pt}
    \end{algorithmic}
\end{algorithm}
Before giving a summary of Algorithm~\ref{alg:safety}, let us define the following {linear program} (LP), where $\optimize\in\{\max,\min\}$:
\begin{align}
    \begin{split}
    \underset{r,\Tr}{\optimize} \quad&
    r(s) + \textstyle\sum_{s'\in\S\setminus\mapq(\target)} \Tr(s'\vert s)V_i(s'),\\
    \text{s.t.}\quad
        &\underline{\Tr}(s'\vert s) \leq \Tr(s'\vert s) \leq \bar{\Tr}(s'\vert s), \\
        & \underline{r}(s) \leq r(s) \leq \bar{r}(s),\\
        & \underline{l}(s) \leq l(s) \leq \bar{l}(s),\\
        & r(s) + l(s) + \textstyle\sum_{s'\in\S\setminus\mapq(\target)} \Tr(s'\vert s) = 1.
    \end{split}\label{eq:LP}
\end{align}
In lines $2$ to $5$ of Algorithm~\ref{alg:safety}, we compute functional lower and upper bounds $(\underline r,\bar r)$ on the \emph{reward} -- the probability of reaching the projected target region $\mapq(\target)\subset\S$ in one time-step starting from the state $s\in\S$.
Analogously, $(\bar l,\underline l)$ represent functional bounds on the \emph{loss} -- the probability of leaving the state space $\S$.
%
In line $7$, the overapproximation $\overline{V}_0$ and underapproximation $\underline{V}_0$
of the \emph{value function} -- the probability of \emph{eventually} reaching the target starting from partition $s$ -- are initialized. 
Until convergence, we repeat the following two steps.
First, for every $s,s'\in\S$, we minimize the LP in \eqref{eq:LP} in line $9$ to find feasible rewards $r$ and transition probabilities $\Tr(s'|s)$ that capture the worst-case behavior of the system.
Then, we update $\overline{V}_0$ and $\underline{V}_0$ via the \emph{Bellman equation} in lines $11$ and $12$.
The while-loop converges when the maximal difference between the over- and under-approximation is less than the predefined closeness bound $\nu$.
After convergence, the under-approximation $\underline{V}_i$ is taken as the lower bound $V_{\text{min}}$ on the value function in line $16$.
The same procedure is repeated to obtain an overapproximation $V_{\text{max}}$ in lines $17$ to $26$, where the LP is maximized in line $19$.
From Proposition~\ref{prop:value} we get that the true latent value $V^*$ lies between the obtained upper and lower bounds.

Following from
Theorem~\ref{thm:error} and Proposition~\ref{prop:value}, 
Theorem~\ref{thm:satProb} gives bounds on the robust satisfaction probability.

\fi

\end{document}